\documentclass[12pt]{article}
\usepackage{amsmath}
\usepackage{graphicx}
\usepackage{enumerate}
\usepackage{url} 

\newcommand{\blind}{1}

\usepackage[normalem]{ulem}

\usepackage{amssymb}
\usepackage{amsmath,amsthm, mathtools,commath}
\usepackage{graphics, color}
\usepackage{booktabs}
\usepackage{siunitx}
\usepackage{arydshln}

\usepackage{graphicx}
\usepackage{epsfig}
\usepackage{makeidx}
\usepackage{makecell}
\usepackage{kotex}
\usepackage{fullpage}
\usepackage{comment}

\usepackage[round]{natbib}
\newcommand{\T}{\top}

\newcommand{\biblist}{\begin{list}{}
{\listparindent 0.0cm \leftmargin 0.50cm \itemindent -0.50 cm
\labelwidth 0 cm \labelsep 0.50 cm
\usecounter{list}}\clubpenalty4000\widowpenalty4000}
\newcommand{\ebiblist}{\end{list}}

\newcommand{\bx}{\bm{x}}

\newcommand{\bX}{\mathbf{X}}

\newcommand{\wh}[1]{\widehat{#1}}

\newcommand{\bbeta}{\boldsymbol{\beta}}
\newcommand{\blambda}{\boldsymbol{\lambda}}
\newcommand{\bomega}{\boldsymbol{\omega}}

\DeclareMathOperator*{\argmin}{arg\,min}

\usepackage{tikz}
\usetikzlibrary{positioning,arrows.meta}
\usepackage{xcolor}

\usepackage{graphicx, graphics, color, subcaption}

\usepackage{amsmath, amssymb, amsfonts, amsthm}
\usepackage{mathtools, mathrsfs, bbm, bm, commath}

\newcommand{\bg}{\widetilde{D}_G}

\newcommand{\what}{\widehat{\bomega}}
\newcommand{\wzero}{\bomega^{(0)}}

\newcommand{\E}{\mathbb{E}}

\newtheorem{theorem}{Theorem}[section]
\newtheorem{lemma}{Lemma}[section]
\newtheorem{proposition}{Proposition}
\newtheorem{corollary}{Corollary}[section]

\newtheorem{assumption}{Assumption}

\theoremstyle{definition}

\theoremstyle{remark}
\newtheorem{remark}{Remark}

\usepackage{float, multirow}

\usepackage{comment}

\usepackage{enumitem} 

\numberwithin{equation}{section}

\allowdisplaybreaks[4]

\begin{document}
\def\spacingset#1{\renewcommand{\baselinestretch}%
{#1}\small\normalsize} 
\spacingset{1}

\if1\blind
{
  \title{\bf  
 Bregman  projection  for calibration estimation in survey sampling 
   }

\author{  Jae Kwang Kim \and Yonghyun Kwon \and Yumou Qiu} 
\date{}
\maketitle
} \fi

\if0\blind
{
 \bigskip
 \bigskip
 \bigskip
 \begin{center}
   {\LARGE\bf 
 Bregman projection for  calibration estimation    
      }
\end{center}
 \medskip
} \fi

\bigskip


\begin{abstract}
Calibration weighting is a fundamental tool in survey sampling for
incorporating auxiliary population information into design-based estimators.
Classical formulations measure distance between calibrated and design weights
on the multiplicative ratio scale. We develop a unified framework based on
Bregman divergence defined directly on the weight vector. The framework
reveals a primal--dual symmetry in which both the weight-space and
multiplier-space optimization problems are themselves Bregman projections, and
the calibrated weights satisfy a generalized Pythagorean decomposition with
respect to the constraint manifold. The resulting estimator is asymptotically
equivalent to a debiased prediction estimator whose regression coefficient
depends explicitly on the Bregman generator, in contrast to the generalized
regression estimator equivalent of classical calibration. Exploiting this
dependence, we identify a contrast-entropy generator that achieves
design-optimality under Poisson sampling. Two extensions are developed: cross-fitted estimation under non-probability sampling, yielding doubly robust inference under standard product-rate conditions; and a regularized extension whose Lagrangian dual produces a Hölder-conjugate penalty for soft balance under high-dimensional auxiliary variables.  Simulations and
an analysis of NOAA's Large Pelagics Intercept Survey illustrate the
framework.
\end{abstract}

\newpage 

\spacingset{1.9} 

\section{Introduction}\label{sec:intro}

Calibration weighting plays a central role in modern survey statistics,
providing a principled mechanism for incorporating auxiliary population
information into design-based estimators to correct for selection bias and
improve efficiency. Given an initial set of design weights, calibration
adjusts these weights to reproduce known totals of auxiliary variables while
remaining close to the original design \citep{fuller2002regression}.
 
The seminal formulation of \citet{deville1992} expressed calibration as the
minimization of a convex distance between calibrated and design weights
subject to linear balancing constraints, encompassing the exponential-tilting
estimator \citep{hainmueller2012}, pseudo empirical likelihood
\citep{wurao2006}, and the generalized regression estimator
\citep{fuller2002regression} as special cases. The framework has become
foundational in finite-population inference \citep{devaud2019} and
model-assisted estimation \citep{breidt2017model}, and parallel reweighting
methods have been developed for covariate balancing in causal inference
\citep{imai2014, chan2016globally, zhao2019}. Most recently,
\citet{kwon2024} introduced a generalized entropy calibration framework based
on the proper scoring rules of \citet{gneiting2007strictly}.
 
These methods share a common formulation: the divergence between calibrated
and initial weights is measured on the multiplicative ratio
$\omega_i / \omega_i^{(0)}$ rather than on the weight vector itself
\citep{csiszar2004}. The geodesic from this construction gives the shortest
path on weight ratios, not on weights themselves, and the class of
representable divergences is correspondingly restricted. This restriction has
consequences that we make precise: the asymptotic regression coefficient in
the equivalent prediction estimator does not depend on the divergence
function, foreclosing efficiency tuning through generator choice, and certain
generators central to design-optimality cannot be expressed at all.
 
In this paper we develop a comprehensive framework for calibration weighting
based on the Bregman divergence \citep{amari2000} defined directly on the
weight vector. The framework reveals an elegant primal--dual structure: both
the weight-space and multiplier-space optimization problems are themselves
Bregman projections, with the convex conjugate $F$ serving as the divergence
in the dual; the calibrated weights satisfy a generalized Pythagorean
decomposition with respect to the constraint manifold; and the equivalent
prediction estimator has a regression coefficient that depends explicitly on
the Bregman generator $G$. This last feature is the structural mechanism
enabling efficiency tuning: by selecting $G$ so that its curvature aligns
with the design variance structure, one obtains an estimator that no
ratio-scale divergence can match. Under Poisson sampling, a unique
contrast-entropy generator achieves design-optimality.
 
The contributions of the paper are as follows. First, we develop the
Bregman framework for calibration weighting that operates directly in weight
space, revealing the calibrated weights as a Bregman projection with a
primal--dual symmetry that reduces an $n$-dimensional constrained
optimization to an unconstrained optimization in $p$ dimensions, and a
generalized Pythagorean decomposition that quantifies the calibration cost.
Second, we show that the resulting estimator is asymptotically equivalent to
a debiased prediction estimator whose regression coefficient depends
explicitly on the generator $G(\cdot)$---a structural feature absent from
the Deville--S\"arndal framework---and identify a contrast-entropy generator
that achieves design-optimality under Poisson sampling. Third, we establish
asymptotic properties under both probability and non-probability sampling:
under known inclusion probabilities, design consistency, asymptotic
normality, and a consistent variance estimator follow under standard
conditions; under unknown inclusion probabilities estimated by cross-fitting
with flexible learners, doubly robust inference holds when the product of
the propensity and outcome approximation errors vanishes faster than
$N^{-1/2}$. Fourth, for high-dimensional auxiliary variables, we develop a regularized extension that replaces exact balance with $\ell_q$-norm tolerance constraints; Lagrangian duality produces a Hölder-conjugate penalty that, for $q = \infty$, performs implicit selection of calibration variables in the spirit of Lasso regularization. 
 
The remainder of the paper is organized as follows.
Section~\ref{sec:basic} introduces the setup and notation.
Section~\ref{sec:method} presents the Bregman framework, its primal--dual
structure, the Pythagorean decomposition, and a positioning relative to
related work in survey sampling, information geometry, and covariate
balancing. Section~\ref{sec:theory} develops the asymptotic theory under
both design-based survey sampling and non-probability sampling with
estimated inclusion probabilities, and characterizes the design-optimal
contrast-entropy generator. Section~\ref{sec:var_select} extends the framework to high-dimensional auxiliary variables. 
Sections~\ref{sec:sim} and~\ref{sec:realdata} report simulation evidence and
an analysis of NOAA's Large Pelagics Intercept Survey.
Section~\ref{sec:discussion} concludes. All technical proofs are relegated
to the supplementary material.

\section{Setting and preliminary}
\label{sec:basic}

Let $\bX$ and $Y$ denote the covariates and response of interest, where $\bX = (X_1, \ldots, X_p)^{\top}$. Let $\{(\bx_{i}, y_{i}): i=1, \ldots, N \}$ be $N$ realized values of $(\bX, Y)$.
Suppose that $y_i$ is subject to missingness and $\bx_i$ are always observed. 
Let $\delta_i$ be the response indicator function such that $y_i$ is observed if and only if $\delta_i=1$. We assume 
that $\delta_i$ follows a Bernoulli distribution given $(\bx_{i}, y_{i})$, and $\pi_i = \mathbb{P}( \delta_i =1 \mid \bx_i, y_i)$ denotes the probability of observing $y_i$. In the probability sampling context, $\{\pi_i\}$ are known for the sample elements.  In the non-probability sampling  context, $\{\pi_i\}$ are unknown and need to be estimated under a propensity score (PS) model. 
From the partially observed data $\{ (\bx_i, \delta_i, \delta_i y_i ); i=1, \ldots, N \}$, we are interested in estimating $\theta = \sum_{i=1}^N y_i$ in both the survey sampling and non-probability sampling settings.

Let $S$ be the index set of sample with $\delta_i=1$. We are interested in using a linear estimator 
$ \widehat{\theta} =  \sum_{i \in S} \omega_i y_i$  
to estimate $\theta$. Let $n$ denote the sample size of $S$. Regarding the conditions on the weights,  we require that the final weights satisfy 
\begin{equation}
\sum_{i \in S} \omega_i \bx_i =  \sum_{i=1}^N \bx_i . 
\label{calib}
\end{equation}
Condition (\ref{calib}), often called the calibration constraint \citep{deville1992} or covariate-balancing constraint \citep{imai2014}, is motivated from a linear regression model:
\begin{equation}
y_i = \bx_i^\top \bbeta + e_i 
\label{reg}
\end{equation}
where $e_i$ satisfies $\E(e_i \mid \bx_i)=0$. If the response mechanism is missing-at-random (MAR), the calibration condition leads to unbiased estimation under the regression model in (\ref{reg}).


To uniquely determine $\omega_i$,  
\cite{deville1992} addresses the problem by minimizing a distance measure between the final calibrated weights and the initial design weights, subject to  (\ref{calib}). Let $\bomega = \{\omega_i: i \in S\}$ and $\bomega^{(0)} = \{\omega_i^{(0)}: i \in S\}$, where $\omega_{i}^{(0)} = \pi_i^{-1}$ in their setting. The objective function of this framework can be expressed as 
\begin{equation}
Q(\bomega \parallel \bomega^{(0)})=\sum_{i\in S}\omega_{i}^{(0)} G \big(\omega_{i} / \omega_{i}^{(0)}\big)
\label{deville}
\end{equation}
where $G(\cdot)\ge0$ is a strongly convex and differentiable function with $g(1)=0$, and $g(\cdot) = G'(\cdot)$ denotes the first order derivative of $G(\cdot)$. 
Let $\widehat{\bomega}_{\rm ds} = (\widehat{\omega}_{{\rm ds}, i}: i \in S) = \operatorname{argmin}_{\bomega} Q(\bomega \parallel \bomega^{(0)})$ subject to the calibration constraint in (\ref{calib}). 
The Deville-Särndal's (DS) calibration estimator of $\theta$ is $\widehat{\theta}_{\rm ds} = \sum_{i \in S} \widehat{\omega}_{{\rm ds}, i} y_i$.
This approach has been widely adopted and has been shown to produce estimators that are asymptotically equivalent to the generalized regression estimator (GREG) estimator $\widehat{\theta}_{\rm greg} = \sum_{i=1}^{N}\bm x_i^\T \widehat {\bm \beta} + \sum_{i \in S} \omega_{i}^{(0)} (y_i - \bm x_i^\T \widehat{\bm \beta})$, where $\widehat{\bm \beta} = \big(\sum_{i \in S} \omega_{i}^{(0)} \bm x_i \bm x_i^\T \big)^{-1}\sum_{i \in S} \omega_{i}^{(0)} \bm x_i y_i$. Although well established, the Deville-Särndal method is just one of several paths to achieving calibration. As discussed in Section \ref{sec:intro},  DS optimization operates on the
ratio $\omega_i/ \omega_i^{(0)}$ rather than directly on the weight vector itself. The geodesic from this projection gives the shortest path on the ratio of weights. However, it does not necessarily imply the shortest path on the weight vector.

\section{Bregman divergence framework}
\label{sec:method}

We now introduce a novel calibration framework that leverages the Bregman divergence as its core distance measure. Let $G(\cdot): \mathcal V \to \mathbb R$ be a prespecified function that is strongly convex and twice-continuously differentiable. Let $g(\omega) = G'(\omega)$. The domain of $G$ is an open interval $\mathcal V = (\nu_1, \nu_2)$ in $\mathbb R$, where $\nu_1 > 0$ and $\nu_2$ is allowed to be $\infty$. For a convex function $G(\cdot)$, define 
\begin{equation}
D_G ( \omega_i \parallel \omega_i^{(0)} ) = G( \omega_i ) - G( \omega_i^{(0)} ) - g( \omega_i^{(0)} ) ( \omega_i - \omega_i^{(0)} )  \label{bd}    
\end{equation}
to be the Bregman divergence of $\omega_i$ evaluated at $\omega_i^{(0)} \in \mathcal{V}$ using $G( \cdot)$ as the generator. The Bregman divergence represents the difference between $G(\omega_i)$ and its tangent line evaluated at $\omega_i^{(0)}$, where the initial weight $\omega_{i}^{(0)}$ takes the form $\omega_i^{(0)} = (n / N) \pi_i^{-1}$ if $\{\pi_i: i \in S\}$ are known or replacing $\pi_i$ by its estimate $\widehat{\pi}_i$ if $\{\pi_i: i \in S\}$ are unknown. Since $G( \cdot)$ is strongly convex, we can establish $D_G ( \omega_i \parallel \omega_i^{(0)} ) \ge 0$ with equality at $\omega_i=\omega_i^{(0)}$. 

Let $\widetilde{D}_{G}(\bomega \parallel \bomega^{(0)}) = \sum_{i\in S} D_G \big( \omega_i \parallel \omega_i^{(0)} \big)$ and $\mathbf{T}_x = (n / N) \sum_{i=1}^N \bx_i$.
We propose to obtain the calibrated weights $\widehat{\bomega} = \{\widehat{\omega}_{i}: i \in S \}$ by minimizing the Bregman divergence measure $\widetilde{D}_{G}(\bomega \parallel \bomega^{(0)})$ subject to a rescaled calibration constraint in (\ref{calib}), where
\begin{equation}
\{\widehat{\omega}_{i}: i \in S\} = \underset{\omega_{i} \in \mathcal V, \, i \in S}{\operatorname{argmin}} \,\,
\widetilde{D}_{G}(\bomega \parallel \bomega^{(0)}) \quad \mbox{subject to} \quad \sum_{i \in S} \omega_i \bx_i = \mathbf{T}_x.
\label{bregman}
\end{equation}
The Bregman calibration (BC) estimator of $\theta$ is 
\begin{equation}\label{bregman-calibration}
    \widehat{\theta}_{\rm BC} = \frac{N}{n} \sum_{i \in S} \widehat\omega_i y_i.
\end{equation}
The Bregman divergence offers a general and theoretically rich foundation for a broad class of calibration methods. This framework is particularly promising due to its elegant algebraic properties, providing a powerful duality principle. 

To solve this constrained minimization problem, the Lagrangian multiplier method is employed. That is, we maximize 
\begin{equation}
\mathcal{L} \big( \bomega, \blambda \big) = - \sum_{i \in S} D_G \big( \omega_i \parallel \omega_i^{(0)} \big) + \blambda^\top \bigg(  \sum_{i \in S} {\omega_i} \bx_i - 
\mathbf{T}_x \bigg)
\label{lagrangian}
\end{equation} 
with respect to $\bomega$, and then minimize this objective function with respect to the Lagrangian multiplier $\blambda$.
Setting $\partial \mathcal{L}/\partial \omega_i =0$ and $\partial \mathcal{L}/\partial \blambda =0$ give the Karush–Kuhn–Tucker (KKT) conditions, which 
yield the expression for the calibration weights as a function of the Lagrange multipliers:
\begin{equation}
\omega_{i}^\star (\blambda) = g^{-1}\{g(\omega_{i}^{(0)})+\bx_{i}^{\top}\blambda\}. 
\label{eq:wgt}
\end{equation}

Because the objective function is strongly convex and the calibration constraints are affine, Slater’s condition is satisfied and strong duality holds,
guaranteeing a unique optimizer if one exists. 
Therefore, $\widehat{\bomega} = \{\widehat{\omega}_{i}: i \in S\}$ is a solution to (\ref{bregman}) if and only if $\widehat{\omega}_{i} = \omega_{i}^\star (\widehat{\blambda}) = g^{-1}\{g(\omega_{i}^{(0)})+\bx_{i}^{\top}\widehat{\blambda}\}$ and $\widehat{\blambda}$ satisfies $\sum_{i \in S} \widehat{\omega}_i \bx_i = \mathbf{T}_x$. In practice, we could also impose $ M_1 \le \wh{\omega}_i \le M_2 $ for some $M_1, M_2\ge 0$. Bregman calibration with range restriction is discussed in Section A.2 of the SM. 

\subsection{Bregman primal--dual structure}\label{subsec:primal-dual}

By plugging (\ref{eq:wgt}) into (\ref{lagrangian}), we obtain the dual objective function: 
\begin{eqnarray} 
\ell ( {\blambda}) = \mathcal{L} \{ \bomega^\star({\blambda}), {\blambda} \} = - \sum_{i \in S} D_G \big( \omega_i^\star ( \blambda) \parallel \omega_i^{(0)} \big) + {\blambda}^\top \bigg( \sum_{i \in S} \omega_i^\star( {\blambda} )
\bx_i - \mathbf{T}_x \bigg).
\label{eq:lagrangian-fun}
\end{eqnarray}
Let $g(\mathcal V) = \{ g(\omega): \omega \in \mathcal V\}$ and $\Lambda = \big\{\bm \lambda: g(\omega_{i}^{(0)}) + \bm x_i^\T \bm \lambda \in g(\mathcal V) \mbox{ \ for all \ } i \in S \big\}$. 
To facilitate the derivation of the dual problem, it is useful to introduce the convex conjugate function 
$F(\nu) = \sup_{\omega \in \mathcal{V}} \, \{ \omega \nu - G (\omega) \} =\nu g^{-1}(\nu)-G\{g^{-1}(\nu)\}$ for $\nu \in g(\mathcal V)$. Using the convex conjugate function, we can establish that  
\begin{equation}
     \ell( {\blambda}) = \sum_{i \in S}  F \big\{ g( \omega_i^{(0)} )  + \bx_i^\top {\blambda}  \big\} -  {\blambda}^\top\mathbf{T}_x + C(\bomega^{(0)})     
     \label{dual}
 \end{equation}
for $\blambda \in \Lambda$, where $C(\bomega^{(0)}) = \sum_{i \in S} \{ G ( \omega_i^{(0)} ) -  g ( \omega_i^{(0)} ) \omega_i^{(0)} \}$ is a function of $\bomega^{(0)}$ only. Note that $F' (\nu) = g^{-1} (\nu)$, where $F'(\cdot)$ denotes the derivative of $F(\cdot)$.
Thus, 
\begin{eqnarray*}
\ell' ( {\blambda} )  
= \sum_{i \in S} F' \big\{ g( \omega_i^{(0)} )  + \bx_i^\top  {\blambda}  \big\} 
\bx_i - \mathbf{T}_x
= \sum_{i \in S} \omega_i^\star({\blambda}) \bx_i - \mathbf{T}_x,
\end{eqnarray*}
and $\ell' ( {\blambda})=0$ leads to  calibration equation. 

Therefore, the  optimization problem for calibration weighting  can be approached from a dual perspective, which often provides a more computationally efficient solution. The primal problem involves finding the optimal weights $\widehat{\bomega}$ by minimizing the Bregman divergence with calibration constraints, an $n$-dimensional optimization task. The dual problem transforms this into an unconstrained optimization problem over the Lagrange multiplier vector $\blambda$, which has a dimensionality $p$ equal to the number of auxiliary variables. This is a significant computational advantage when the sample size $n$ is much larger than $p$. 

\begin{table}[ht]
\centering
\begin{tabular}{cccc}
\hline
Generalized Entropy                                             & $G(\omega)$                 & $F(\nu)$        \\ \hline
Squared loss                                       & $\omega^2/2$                  & $\nu^2/2$              \\  
Kullback-Leibler & $\omega \log (\omega)$ & $\exp ( \nu -1)$ \\ 
Shifted KL & $(\omega-1) \{ \log (\omega-1)- 1\}$ & $
\nu + \exp ( \nu )$  \\
Empirical likelihood & $- \log ( \omega)$ & $-1-\log ( - \nu)$   \\
Squared Hellinger & $( \sqrt{\omega} -1 )^2$ & $\nu/(1 - \nu)$  \\
R\'enyi entropy $(\alpha \neq 0, -1)$ 
& $ (\alpha+1)^{-1} \omega^{\alpha + 1}$   & $\alpha (\alpha+1)^{-1} \nu^{\frac{\alpha+1}{\alpha}}$ \\ 
\hline
\end{tabular}
\caption{Examples of generalized entropies, $G(\omega)$ and the corresponding convex conjugate function $F(\nu)$. } 
\label{tab:15-1}
\end{table}

A deeper examination of the dual problem reveals a profound mathematical structure. By the definition of the convex conjugate $F$ (Legendre transformation) of $G$, 
we have $F ( \nu) + G(\omega) \ge  \omega \nu$ 
holds for any $\omega \in \mathcal{V}$ and $\nu \in g(\mathcal{V})$. The equality holds if and only if $\omega$ and $\nu$ satisfy the first-order condition 
\begin{equation} \label{eq:theta-eta-relation}
 \nu = G' (\omega) \mbox{ \ and \ }
 \omega = F' (\nu)
\end{equation} 
for the optimization problem $F(\nu)= \sup_{\omega \in \mathcal{V}} \{ \omega \nu - G (\omega) \}$, where the second equality in (\ref{eq:theta-eta-relation}) is from the symmetry of the convex conjugate.


\begin{lemma}\label{lem:dual-bregman}
Suppose that there exists a $\widehat{\blambda} \in \Lambda$ such that $\ell'(\widehat{\blambda}) = 0$. Then, $\widehat{\blambda}$ is the unique minimum of $ \ell({\blambda})$ for $\blambda \in \Lambda$ and the solution to the primal problem in (\ref{bregman}) exists and unique where $\widehat{\omega}_{i} = g^{-1}\{g(\omega_{i}^{(0)})+\bx_{i}^{\top}\widehat{\blambda}\}$. Furthermore, the Lagrangian dual objective function $\ell ( {\blambda})$ in (\ref{eq:lagrangian-fun}) can be expressed via a Bregman divergence as
\begin{equation}
  \ell(\boldsymbol{\lambda}) - \ell(\widehat{\boldsymbol{\lambda}})
  = \sum_{i\in S} D_F\!\bigl(\nu_i(\boldsymbol{\lambda})
    \,\|\, \nu_i(\widehat{\boldsymbol{\lambda}})\bigr) \;\geq\; 0,
\label{eq:3-10}
\end{equation}
where $\nu_i(\blambda)=g(\omega_i^{(0)})+\bx_i^\top\blambda$ and the equality holds if and only if $\nu_i(\boldsymbol{\lambda}) = \nu_i(\widehat{\boldsymbol{\lambda}})$ for all $i$.
\end{lemma}

By Lemma~\ref{lem:dual-bregman}, the dual objective function, $\ell(\boldsymbol{\lambda})$, can be expressed as a sum of Bregman divergences of the convex conjugate function $F$. The result implies that the dual problem is fundamentally a Bregman projection problem in its own right, but operating in the space of the Lagrange multipliers and with the convex conjugate function as the divergence measure.

\begin{figure} 

\begin{center} 
    \begin{tikzpicture}[>=stealth, node distance=5cm, thick]
  \node[draw, rounded corners, minimum width=3cm, minimum height=1.2cm,
        fill=blue!10, align=center] (primal)
    {Primal space \\[4pt] Weights $\omega$\\ Divergence $D_G(\cdot \parallel \cdot)$};

  \node[draw, rounded corners, minimum width=3cm, minimum height=1.2cm,
        fill=red!10, align=center, right=of primal] (dual)
    {Dual space \\[4pt] Multipliers $\blambda$ \\ Parameters $\nu$\\ Divergence $D_F(\cdot \parallel \cdot)$};

  \draw[->, thick] (primal) -- node[midway, above, align=center]
    {$\nu= g(\omega)=G'(\omega)$ \\ (calibration link)} (dual);
  \draw[->, thick] (dual) -- node[midway, below, align=center]
    {$\omega =  g^{-1}(\nu)= F'( \nu)$ \\ (inverse calibration link)} (primal);

  \end{tikzpicture}
\caption{Primal--dual structure of Bregman calibration weighting. The calibration link $g = G'$ maps the weight space to the natural parameter space; the inverse calibration link $g^{-1} = F'$ maps back. Both spaces carry their own Bregman divergence, and the two optimization problems---minimizing $D_G$ in the primal and $D_F$ in the dual---are coupled through this link.} 
\label{fig:1}
\end{center}
 
\end{figure}
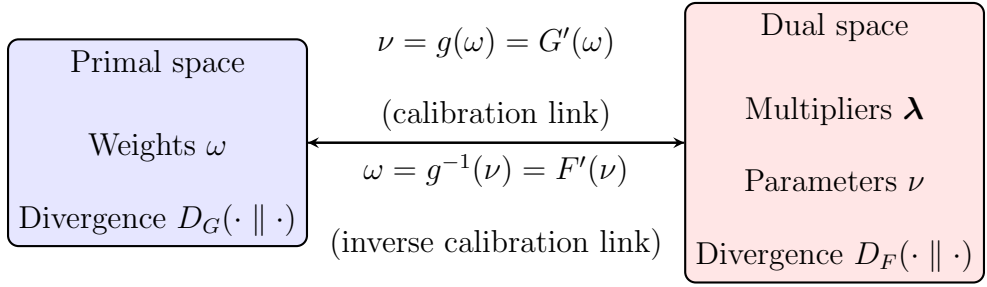

The map that connects the primal and dual spaces is the derivative $g = G'$, which we call the \emph{calibration link function}. Its inverse $g^{-1} = F'$ is the \emph{inverse calibration link}. Figure~\ref{fig:1} illustrates this dual structure involving the two coupled coordinate systems $\omega$ and $\nu$, which are connected by the transformation $\nu = g(\omega)$ and $\omega = g^{-1}(\nu)$.
The calibration link is one-to-one and differentiable, and it establishes a precise correspondence between the weight space and the dual (natural parameter) space. The primal problem minimizes the Bregman divergence $D_G$ in the weight space; the dual problem minimizes the conjugate divergence $D_F$ in the natural parameter space.

This terminology is motivated by a structural parallel with generalized linear models (GLMs), where the conditional mean $\mathbb{E}(Y_i \mid \bX_i = \bx_i) = b'(\nu_i)$ is related to the linear predictor $\nu_i = \bx_i^{\top}\boldsymbol{\beta}$ through the cumulant function $b(\cdot)$, and the canonical link $(b')^{-1}(\cdot)$ maps mean parameters to natural parameters. In our context, the calibration link operates on the weight parameter~$\omega_i$ rather than a conditional mean, but the algebraic structure is identical: the convex conjugate $F$ plays the role of the cumulant function, and the weight map in \eqref{eq:wgt}, $\omega_i^{\star}(\boldsymbol{\lambda}) 
= g^{-1}\!\bigl\{g(\omega_i^{(0)}) + \bx_i^{\top}\boldsymbol{\lambda}\bigr\}
= F'\!\bigl\{g(\omega_i^{(0)}) + \bx_i^{\top}\boldsymbol{\lambda}\bigr\}$,
takes the form of a GLM with canonical link $g$, inverse link $g^{-1} = F'$, and unit-specific offset $g(\omega_i^{(0)})$. The calibrated weight is obtained by shifting the natural parameter of the baseline weight by $\bx_i^{\top}\boldsymbol{\lambda}$ and mapping back to the weight space through the inverse link. 

\subsection{Pythagorean decomposition}

One of the most elegant and powerful properties of Bregman divergence is the Pythagorean theorem. It states that for any vector of weights $\bomega$ that satisfy the calibration constraints, its Bregman divergence from the initial weights can be decomposed into two components: the divergence from $\bomega$ to the optimal solution $\widehat{\bomega}$ to (\ref{bregman}) and the divergence from $\widehat{\bomega}$ to the initial weights. This decomposition is analogues to the sum-of-square decomposition for linear models in linear projection spaces. Our formulation extends such orthogonal decompositions to the manifold of probability weights.

\begin{theorem}[Pythagorean equality]\label{theorem:4.1}
Let $\mathcal{C} = \big\{\bomega = (\omega_1, \ldots, \omega_n)^{\T} \in \mathbb{R}^{n}: 
\sum_{i \in S} \omega_i \bx_i = \mathbf{T}_x \mbox{ and } \omega_i > 0 \mbox{ for all } i\}$ be the space satisfying the calibration constraints, where $n$ is the sample size of $S$. For any element $\bomega\in\mathcal{C}$, we have
\begin{equation}
\bg (\bomega \parallel \bomega^{(0)})=\bg (\bomega \parallel \widehat{\bomega})+ \bg (\widehat{\bomega} \parallel \bomega^{(0)}), 
\label{pytha}
\end{equation}
where 
\begin{equation}
\widehat{\bomega} = \omega_{i}^\star (\widehat{\blambda})=g^{-1}\big(g(\omega_{i}^{(0)})+\bx_{i}^{\top}\widehat{\blambda}\big)
\label{eq:wgt-1}
\end{equation}
and $\widehat{\blambda}$  is the unique minimizer of $ \ell({\blambda})$ for $\blambda \in \Lambda$.  
\end{theorem}

This theorem provides a powerful geometric interpretation: the optimal weights $\widehat{\bomega}$ are the ``Bregman projection'' of the initial weights $\bomega^{(0)}$ onto the constraint space $\mathcal{C}$. The Pythagorean equality in (\ref{pytha}) implies that 
$$\bg (\bomega \parallel \bomega^{(0)}) \ge \bg (\widehat{\bomega} \parallel \bomega^{(0)}), \ \ \ \mbox{for any} \ \bomega \in \mathcal{C},$$
which shows that $\widehat{\bomega}$ is the global minimum.  In essence, $\widehat{\bomega}$ in (\ref{eq:wgt-1}) is the closest point in the constraint set to the initial weights under the Bregman divergence generated by
$\widetilde G(\bomega) = \sum_{i \in S}  G(\omega_i).$ That is, 
among all feasible weights, \(\what\) is uniquely closest to \(\wzero\) in the geometry induced by \(G\).
The nonnegative scalar \(\bg(\what \parallel \wzero)\) quantifies the unavoidable distortion needed to
achieve balance and can be reported as the value for \emph{calibration cost}. 

\subsection{Related work and structural distinctions}
\label{sec:related-work}

The proposed framework sits at the intersection of three research
traditions: calibration in survey sampling, Bregman methods in information
geometry, and covariate balancing in causal inference. We outline the key
connections and structural distinctions from each.

\paragraph{Calibration in survey sampling.}
The dominant framework is the Deville--S\"arndal (DS) construction
\citep{deville1992}, which minimizes a divergence on weight ratios
$\omega_i/\omega_i^{(0)}$ and subsumes pseudo empirical likelihood
\citep{wurao2006}, entropy balancing \citep{hainmueller2012entropy}, and
the generalized regression estimator \citep{fuller2002regression} as special
cases; \citet{devaud2019} and \citet{breidt2017model} provide recent
overviews. The
Bregman framework, motivated from \citep{kwon2024},  operates directly in weight space, equating
$g(\omega_i) - g(\omega_i^{(0)}) = \mathbf{x}_i^\top \bm{\lambda}$ rather than
a function of the ratio, and coincides with DS only when $g$ is affine in
$\log$---that is, only for the exponential-tilting generator. For every
other generator the resulting weights, asymptotic regression coefficients
(Theorem~\ref{thm:CLT-design}), and efficiency properties differ, and the
contrast-entropy generator that achieves design-optimality under Poisson
sampling (Section~\ref{sec:design}) cannot be expressed in DS form.
Relative to GEC, the Bregman framing makes explicit the primal--dual
symmetry in which both the weight-space and multiplier-space problems are
themselves Bregman projections, and supports the doubly robust extension via
cross-fitting in Section~\ref{sec:non-prob-sampling}.

\paragraph{Covariate balancing in causal inference.}
A parallel literature has developed reweighting methods for causal inference
under balancing constraints, including the covariate balancing propensity
score \citep{imai2014}, empirical balancing calibration weighting
\citep{chan2016globally}, tailored loss functions \citep{zhao2019},
and approximately balancing weights \citep{wang2020b}. Several of
these are recovered as special cases of Bregman calibration with specific
generators. The structural distinctions are fourfold: the inferential target
is the finite-population mean under a sampling design rather than a treatment
effect under a counterfactual model; prior work commits to a single
divergence (empirical likelihood, exponential tilting, or a tailored loss)
whereas the Bregman framework treats the divergence as a tunable design
parameter; the generator-dependent asymptotic regression coefficient
(Theorem~\ref{thm:CLT-design}) and resulting efficiency-tuning mechanism do not
appear in this literature; and design-optimality has no natural analog in
causal-inference settings where sampling is not the source of randomization.
The cross-fitting machinery for doubly robust estimation in
Section~\ref{sec:non-prob-sampling} follows the standard pattern in this
literature; its novelty here lies in its embedding within the broader
Bregman framework rather than in the cross-fitting mechanism itself.


\section{Asymptotic properties}
\label{sec:theory}

To examine the asymptotic properties of the proposed Bregman calibration estimator, we consider an increasing sequence of $n$ and $N$.
Throughout this section, we present the asymptotic results on the mean scale. Namely, the target parameter is the finite-population mean $\mu = N^{-1}\sum_{i=1}^N y_i$ in both the survey sampling and non-probability sampling settings. 
We assume that the initial weights are normalized so that $c_1 < \omega_i^{(0)} < c_2$ for positive constants $c_1, c_2$ and all $i \in S$. 

Let $\bar{\bm{X}}_N = N^{-1}\sum_{i=1}^N \bx_i$ be the mean of covariates. For $\blambda\in\mathbb{R}^p$, recall that $\nu_i(\blambda) = g(\omega_i^{(0)}) + \bx_i^\top\blambda$, $\omega_i^\star(\blambda) := F'\{\nu_i(\blambda)\}$, and $\widehat{\blambda}$ is the solution to the mean-scale calibration equation $n^{-1}\sum_{i\in S}\omega_i^\star(\blambda) \bx_i = \bar{\bm{X}}_N$,
which is equivalent to the calibration equation in \eqref{bregman}. The Bregman calibration estimator (BCE) of $\mu$ is
\begin{equation}
\widehat{\mu}_{\rm BC} := \frac{1}{n}\sum_{i\in S}\omega_i^\star(\widehat{\blambda})\,y_i. \label{eq:BCE}
\end{equation}

Let $\blambda_0$ be the solution to the population equation
\[
\mathbb{E}\{\ell'(\blambda)\} = \sum_{i=1}^N \pi_i\,F'\!\big\{g(\omega_i^{(0)}) + \bx_i^\top\blambda\big\}\bx_i - \mathbf{T}_x = 0.
\]
Let $\omega_i^\star := \omega_i^\star(\blambda_0) = F'\{\nu_i(\blambda_0)\}$ and $q_i := F''\{\nu_i(\blambda_0)\}$.
Let $A_n=n^{-1}\sum_{i\in S} q_i\bx_i\bx_i^\top$ and 
\begin{equation*}
\widetilde{\bbeta}_g^\star = \Bigl(\sum_{i\in S} q_i \bx_i \bx_i^\top\Bigr)^{-1}\sum_{i\in S} q_i \bx_i y_i. 
\end{equation*}
Let $\|\cdot\|$ denote Euclidean norm for vectors and spectral norm for matrices when there is no confusion. 
In the following theorem, we assume $\blambda_0$ exists and unique. This condition is verified under probability sampling and non-probability sampling settings in Sections \ref{sec:design} and \ref{sec:non-prob-sampling}, respectively.

\begin{lemma}[Expansion of BCE]\label{thm:expansion}
For a given sample $\{(\bx_i, \delta_iy_i, \delta_i): i = 1, \ldots, N\}$, suppose that the solutions $\widehat{\blambda}, \blambda_0 \in \Lambda$ to $\ell'(\widehat{\blambda}) = 0$ and $\mathbb{E}\{\ell'(\blambda_0)\} = 0$ exist and $\|\widehat{\blambda} - \blambda_0\| \leq \epsilon$ for a sufficiently small $\epsilon$. 
If $c_1 < \omega_i^\star < c_2$ for all $i \in S$, $\|A_n^{-1}\| < c_3$, $n^{-1}\sum_{i\in S} \|\bx_i\|^2|y_i| < c_3$, $n^{-1}\sum_{i\in S} \|\bx_i\|^3 < c_3$, and $n^{-1}\sum_{i\in S} \|\bx_i\||y_i| < c_3$ for positive constants $c_1, c_2, c_3$, the Bregman calibration estimator in~\eqref{eq:BCE} admits the expansion
\begin{equation}
\widehat{\mu}_{\rm BC} = \bar{\bm{X}}_N^\top\widetilde{\bbeta}_g^\star + \frac{1}{n}\sum_{i\in S}\omega_i^\star\bigl(y_i - \bx_i^\top\widetilde{\bbeta}_g^\star\bigr) + O(\epsilon^2). \label{eq:expansion}
\end{equation}
\end{lemma}

Lemma \ref{thm:expansion} establishes the expansion of the proposed BCE $\widehat{\mu}_{\rm BC}$ for a given sample under the required conditions on the sample. Although it isn't the asymptotic result as $n, N \to \infty$, the asymptotic expansion of $\widehat{\mu}_{\rm BC}$ can be established under either probability sampling or non-probability sampling setting using this lemma. It implies the asymptotic equivalence between the proposed BCE $\widehat{\mu}_{\rm BC}$ and the debiased prediction estimator
\begin{equation}
\widehat{\mu}_{\rm DP} = \frac{1}{N}\sum_{i=1}^N \bx_i^\top\widehat{\bbeta}_g^\star + \frac{1}{n}\sum_{i\in S} \omega_i^\star(\widehat{\blambda}) (y_i - \bx_i^\top\widehat{\bbeta}_g^\star)
\label{eq:DP}
\end{equation}
if $\widehat{\blambda}\to\blambda_0$ as $n, N \to \infty$,
where $\widehat{\bbeta}_g^\star = \bigl(\sum_{i\in S}\widehat{q}_i\,\bx_i \bx_i^\top\bigr)^{-1}\sum_{i\in S}\widehat{q}_i\,\bx_i y_i$ with $\widehat{q}_i = 1/g'(\omega_i^\star(\widehat{\blambda}))$ is a plug-in estimate of $\widetilde{\bbeta}_g^\star$.
Notably, this equivalence holds without any model assumption on $Y$ or any assumption on the distribution of $\delta_i$. In the following two subsections, we verify the conditions of Lemma \ref{thm:expansion} and show that the proposed estimator is consistent for the population mean under probability sampling and non-probability sampling settings. 

\subsection{Design-based survey sampling setting}\label{sec:design}

The key to show (\ref{eq:expansion}) is to demonstrate the existence of the solution $\widehat{\blambda}$ to $\ell'(\widehat{\blambda}) = 0$ and derive the convergence rate of $\widehat{\blambda}$. In the design-based survey sampling setting, we treat $\bx_i$ and $y_i$ as fixed quantities and $\delta_i$ as random, where the first-order inclusion probability $\pi_i = \mathbb{P}(\delta_i = 1)$ of unit~$i\in S$ is known. In this setting, we use $\omega_i^{(0)} = (n/N)\pi_i^{-1}$ as the initial weight. 

Let $\pi_{ij} = \mathbb{P}(\delta_i = \delta_j = 1)$ be the joint inclusion probability of units $i$ and $j$ and $\Delta_{ij} = \pi_{ij} - \pi_i \pi_j$. Let $\bm \Sigma = \underset{N \to \infty}{\lim} \sum_{i = 1}^{N} \bm x_i {\bm x_i}^{\T} / N$ and $\bm \Gamma(\bm \lambda) = \underset{N \to \infty}{\lim} \sum_{i = 1}^{N} F''(g(\omega_i^{(0)}) + \bm \lambda^\T \bm x_i) \bm x_i {\bm x_i}^{\T} / N$. Let $n_0 = {\mathbb E}(n) = \sum_{i = 1}^{N} \pi_i$. The following lemma shows that $\widehat{\blambda} \xrightarrow{p} \bm 0$.

\begin{lemma}\label{lem:lambda-zero}
Assume that $c_1 < n_0 / (N \pi_i) < c_2$ for $i = 1, \ldots, N$ and positive constants $c_1, c_2 \in \mathcal{V}$, $\limsup_{N \to \infty} N^2n_0^{-1} \max_{i \neq j} \abs{\Delta_{ij}} < \infty$, $\bm \Sigma$ exists and positive definite, the average 4th moment of $(y_i, \bm x_i^\T)$ is finite such that $\underset{N \to \infty}{\limsup} \sum_{i = 1}^N \norm{(y_i, \bm x_i^\T)}^4 / N < \infty$, and $\bm \Gamma(\bm \lambda)$ exists in a neighborhood around $\bm \lambda_0= \bm 0$. Then, the solution $\widehat{\blambda}$ to $\ell'(\blambda) = 0$ exists and is unique with probability approaching 1 ($w.p.a.1$), and $\|\widehat{\blambda}\|=O_p(n_0^{-1/2})$.
\end{lemma}

The following theorem presents the asymptotic normality of the proposed BCE under the case of probability sampling in which $\{\pi_i: i \in S\}$ are known.  

\begin{theorem}\label{thm:CLT-design}
Under the assumptions in Lemma~\ref{lem:lambda-zero}, the proposed Bregman calibration estimator $\widehat{\mu}_{\rm BC} = n^{-1} \sum_{i\in S}\widehat{\omega}_i y_i$ satisfies
\begin{equation}
\widehat{\mu}_{\rm BC} = \frac{1}{N}\sum_{i=1}^N \bx_i^\top\widetilde{\bbeta}_g^{(0)} + \frac{1}{N}\sum_{i\in S} \frac{1}{\pi_i} \big(y_i - \bx_i^\top\widetilde{\bbeta}_g^{(0)}\big) + o_p(n_0^{-1/2}), \label{eq:BC-DP-equiv}
\end{equation}
where $\widetilde{\bbeta}_g^{(0)} = \big\{\sum_{i = 1}^{N} \pi_i \bx_i \bx_i^\top / g'(\omega_i^{(0)}) \big\}^{-1} \big\{ \sum_{i = 1}^{N} \pi_i \bx_i y_i / g'(\omega_i^{(0)}) \big\}$.
Furthermore, under additional regularity conditions D1 and D2 on the sampling design in the SM, $\widehat{\mu}_{\rm BC}$ satisfies
\[
{\mathbb V}(\widehat{\mu}_{\rm BC})^{-1/2}\bigl(\widehat{\mu}_{\rm BC} - \mu\bigr) \xrightarrow{d} N(0,1) \quad \mbox{as } n_0, N \to \infty,
\]
where ${\mathbb V}(\widehat{\mu}_{\rm BC}) = \big[ N^{-2}\sum_{i,j=1}^{N}(\pi_{ij} / \pi_i\pi_j - 1)\,\{y_i - \bx_i^\top\widetilde{\bbeta}_g^{(0)} \} \{y_j - \bx_j^\top\widetilde{\bbeta}_g^{(0)} \} \big] \{1+o(1)\}$. 
\end{theorem}

A critical distinction of this framework from the traditional Deville--S\"arndal method is that the regression coefficient $\widetilde{\bbeta}_g^{(0)}$ explicitly depends on the convex function $G(\cdot)$ via $1/g'(\omega_i^{(0)})$, where $g'(\omega) = dg(\omega)/d\omega$. In contrast, the asymptotic GREG estimator in the Deville--S\"arndal framework is independent of the choice of $G(\cdot)$. This dependency can be utilized to improve the efficiency of the calibration estimator.

The variance ${\mathbb V}(\widehat{\mu}_{\rm BC})$ of the proposed estimator can be estimated by
\begin{equation}
\widehat{\mathbb V}(\widehat{\mu}_{\rm BC}) = \frac{1}{N^2}\sum_{i,j\in S}\frac{\pi_{ij}-\pi_i\pi_j}{\pi_{ij}}\,\frac{y_i - \bx_i^\top\widehat{\bbeta}_g^\star}{\pi_i}\,\frac{y_j - \bx_j^\top\widehat{\bbeta}_g^\star}{\pi_j}, \label{eq:var-est-design}
\end{equation}
where $\widehat{\bbeta}_g^\star = \bigl(\sum_{i\in S}\widehat{q}_i\,\bx_i\bx_i^\top\bigr)^{-1}\sum_{i\in S}\widehat{q}_i\,\bx_iy_i$ with $\widehat{q}_i = 1/g'(\widehat{\omega}_i)$. 
By a direct adaptation of the argument in \cite{robinson1983asymptotic} for standard design-based variance estimation of regression estimators, we can show the ratio consistency $\widehat{\mathbb V}(\widehat{\mu}_{\rm BC})/{\mathbb V}(\widehat{\mu}_{\rm BC})\xrightarrow{p} 1$.

Under Poisson sampling, where sampling units are selected independently, the design-optimal regression estimator \citep{montanari1987} is obtained with a specific regression coefficient, $\widetilde{\bbeta}_{\mathrm{opt}}^{\star}$, that depends on the variance structure of the design. The optimal coefficient is given by $\widetilde{\bbeta}_{\mathrm{opt}}^{\star} = \bigl(\sum_{i\in S} q_i \bx_i \bx_i^\top\bigr)^{-1}\sum_{i\in S} q_i \bx_i y_i$, where $q_i = d_i^2 - d_i$ and $d_i = \pi_i^{-1}$. To achieve this design-optimal estimation within the Bregman divergence framework, a specific entropy function $G(\cdot)$ must be chosen such that its derivative satisfies the condition:
\begin{equation}
1/g'(d_i) = d_i^2 - d_i. \label{eq:design-opt-cond}
\end{equation}
The condition~\eqref{eq:design-opt-cond} for design-optimality is satisfied by the contrast-entropy function:
\begin{equation}
G(\omega) = (\omega-1)\log(\omega-1) - \omega\log\omega. \label{eq:ce}
\end{equation}
The existence of this function confirms that the proposed framework is not only a theoretically rich alternative to traditional calibration but can also be used to construct a statistically optimal estimator for the specific case of Poisson sampling.

\subsection{Non-probability sampling with estimated inclusion probabilities}
\label{sec:non-prob-sampling}
 
In this section we consider non-probability sampling, in which units enter
the sample through a self-selection or convenience mechanism rather than a
controlled probability design. Such samples arise in opt-in web panels,
voluntary surveys, administrative data registries, and online crowdsourcing
platforms, and have become an increasingly common supplement to or
substitute for traditional probability samples
\citep{chen2020doubly,wu2022statistical}. The inclusion
probabilities $\{\pi_i\}$ are unknown and must be estimated.
 
We adopt nonparametric estimation of $\pi(x) = P(\delta = 1 \mid X = x)$ via
probabilistic classification using a flexible learner on the observed data
$\{(\mathbf{x}_i, \delta_i)\}_{i=1}^N$---for example, random forests,
boosting, or neural networks. To eliminate overfitting bias and accommodate
general classification learners, we adopt the following cross-fitting
procedure. Fix $K \ge 2$ and draw independent fold labels $\kappa_i \in
\{1, \ldots, K\}$ uniformly for each $i \in \{1, \ldots, N\}$. Let $U^{(k)}
:= \{i : \kappa_i = k\}$. For each fold $k$: (i) estimate
$\widehat{\pi}^{(-k)}(\cdot)$ using only $(\mathbf{x}_i, \delta_i)$ for $i
\notin U^{(k)}$; (ii) define $\widehat{\pi}_i^{(-)} :=
\widehat{\pi}^{(-k)}(\mathbf{x}_i)$ for $i \in U^{(k)}$. The cross-fitted
initial weights for sampled units are
\begin{equation}
\label{eq:cf-initial-weights}
\widehat{\omega}_i^{(0)} = \frac{n}{N} \cdot \frac{1}{\widehat{\pi}_i^{(-)}}
\qquad \text{for } i \in S.
\end{equation}
 
We treat $X$ and $Y$ as random variables and assume the ignorable selection
mechanism $\delta \perp Y \mid X$, which is the analog of missing-at-random
in the missing-data literature and of strong ignorability in causal
inference. Let $m(x) = \mathrm{E}(Y \mid X = x)$ denote the outcome
regression function, satisfying
\begin{equation}
\label{eq:OR-model}
Y = m(X) + \varepsilon, \qquad \mathrm{E}(\varepsilon \mid X) = 0.
\end{equation}
Assume the observed data $\{(\mathbf{x}_i^\top, \delta_i y_i,
\delta_i)\}_{i=1}^N$ are independent and identically distributed and that
$n/N \to c_0 \in (0, 1)$. The Bregman calibration estimator of the mean
$\mu$ with estimated inclusion probabilities is
\begin{equation}
\label{eq:BCE-NPS}
\widehat{\mu}_{\mathrm{BC}, \widehat{\pi}}
= \frac{1}{n} \sum_{i \in S}
\omega_i^{\star}(\widehat{\bm{\lambda}}, \widehat{\pi}) \, y_i,
\end{equation}
where $\widehat{\omega}_i = \omega_i^{\star}(\widehat{\bm{\lambda}},
\widehat{\pi}) = g^{-1}\{g(\widehat{\omega}_i^{(0)}) + \mathbf{x}_i^\top
\widehat{\bm{\lambda}}\}$ are the calibration weights from \eqref{bregman}
using the cross-fitted estimated inclusion probabilities as the initial
weights $\omega_i^{(0)} = \widehat{\omega}_i^{(0)}$. Let
$\widehat{\mu}_{\mathrm{BC}, \pi}$ denote the oracle BCE using the true
inclusion probabilities as the initial weights and
\[
\bbeta_g^{(0)} = \big[ \mathbb{E} \big\{\pi_i \bx_i \bx_i^\top / g'(c_0 d_i) \big\} \big]^{-1} \mathbb{E}\big\{\pi_i \bx_i y_i / g'(c_0 d_i) \big\}
\]
be limit of $\widetilde{\bbeta}_g^{(0)}$, 
where $d_i = \pi_i^{-1}$. Define the approximation error of weighted linear projection:
\begin{equation}
r(\bx) := m(\bx) - \bx^\top\bbeta_g^{(0)}. \label{eq:approx-error}
\end{equation}
The residual from the weighted regression can be decomposed as $e_i^g = y_i - \bx_i^\top\bbeta_g^{(0)} = \varepsilon_i + r(\bx_i)$, where $\varepsilon_i$ is the irreducible noise, orthogonal to $\bx_i$ by construction, while $r(\bx_i)$ measures how well the calibration model $\bx^\top\bbeta_g^{(0)}$ approximates the true conditional mean $m(\bx)$.

For a function $h(\bx)$ of $\bx$, let $\|h(\bx)\|_{L_2} = \bigl\{N^{-1}\sum_{i=1}^N h^2(\bx_i)\bigr\}^{1/2}$ be the Euclidean norm of $(h(\bx_1),\ldots,h(\bx_N))$. We impose the following conditions.

\begin{assumption}[Moments]\label{ass:moments}
$(\bX, Y)$ is sub-Gaussian distributed, $\mathbb E(\bm x_i \bm x_i^\top)$ is positive definite.
\end{assumption}

\begin{assumption}[Positivity]\label{ass:positivity}
There exist constants $0<\pi_{\min}\le\pi_{\max}<1$ such that $\pi_{\min}\le\pi(\bx)\le\pi_{\max}$ a.s., and $\pi_{\min}\le\widehat{\pi}_i^{(-)}\le\pi_{\max}$ for all~$i$, w.p.a.1.
\end{assumption}

\begin{assumption}[Out-of-fold honesty]\label{ass:honesty}
For each fold~$k$, $\widehat{\pi}^{(-k)}(\cdot)$ is measurable with respect to $\mathcal{T}_k := \sigma\{(\bx_i,\delta_i):i\notin U^{(k)}\}$ and is independent of $\{(\bx_i, y_i,\delta_i):i\in U^{(k)}\}$. 
\end{assumption}

\begin{assumption}[Estimation error]\label{ass:est-error}
Let $d(\bx) = \pi(\bx)^{-1}$ and $\widehat{d}_i^{(-)} = (\widehat{\pi}_i^{(-)})^{-1}$. Assume that (a) $\|\widehat{d}^{(-)}(\bx)-d(\bx)\|_{L_2} = o_p(1/\log(N))$; and (b) $\sqrt{N}\,\|\widehat{d}^{(-)}(\bx)-d(\bx)\|_{L_2}\cdot\|r(\bx)\|_{L_2} = o_p(1)$.
\end{assumption}

Assumption~\ref{ass:moments} is a mild standard condition on the distributions of $\bX$ and $Y$. Assumption~\ref{ass:positivity} is the standard overlap (positivity) condition in the missing-data and causal-inference literatures. Assumption~\ref{ass:honesty} formalizes the key benefit of cross-fitting: since $\widehat{\pi}^{(-k)}(\cdot)$ is trained only on out-of-fold data, it is conditionally independent of the response and sampling indicators $(y_i,\delta_i)$ in fold~$k$, given the covariates $\bx_i$. 
This independence ensures that the cross-term involving $\varepsilon_i$ has conditional mean zero, eliminating a potential overfitting bias.

Assumption~\ref{ass:est-error} is the most substantive condition. Part~(a) requires only $L_2$-consistency of the cross-fitted propensity estimator, which is satisfied by virtually every reasonable propensity estimator under the positivity condition. Part~(b) is a product-rate condition whose doubly robust structure is made transparent by the following remark.

\begin{remark}[Doubly robust interpretation of Assumption~\ref{ass:est-error}]\label{rem:DR}
(i) If $m(\bx) = \bx^\top\bbeta_0$ (outcome model correct), then $r(\bx)=0$ and~\ref{ass:est-error}(b) holds trivially.
(ii) If $\pi(x)$ is correctly specified at the parametric $\sqrt{N}$-rate, linearization yields $\sqrt{N}$-consistency without invoking~\ref{ass:est-error}(b).
(iii) When both models are estimated nonparametrically at rates $N^{-a}$ and $N^{-b}$, condition~\ref{ass:est-error}(b) holds whenever $a+b>1/2$.
\end{remark}

The following lemma extends Lemma~\ref{lem:lambda-zero} to estimated propensities.

\begin{lemma}\label{lem:lambda-zero-est}
Suppose $G(\cdot)$ is strongly convex and $F(\cdot)$ is third continuously differentiable. 
Under Assumptions~\ref{ass:moments}--\ref{ass:est-error}, 
the solution $\widehat{\blambda}$ to the dual problem of the Bregman calibration with estimated baseline weights $\widehat{\omega}_i^{(0)} = (n/N)\widehat{d}_i^{(-)}$ exists and is unique $w.p.a.1$, and $\widehat{\blambda}\xrightarrow{p} \mathbf{0}$.
\end{lemma}

We now state the main asymptotic result under the non-probability sampling setting. The following theorem establishes the asymptotic normality of $\widehat{\mu}_{{\rm BC},\widehat{\pi}}$.

\begin{theorem}\label{thm:CLT-missing}
Under the conditions in Lemma~\ref{lem:lambda-zero-est}, for the BCE $\widehat{\mu}_{{\rm BC},\widehat{\pi}}$ with estimated-propensity,
we have $\widehat{\mu}_{{\rm BC},\widehat{\pi}} - \widehat{\mu}_{{\rm BC},\pi} = o_p(N^{-1/2})$, and it inherits the oracle expansion:
\begin{equation}
\widehat{\mu}_{{\rm BC},\widehat{\pi}} = \bar{\bm{X}}_N^\top \bbeta_g^{(0)} + \frac{1}{N}\sum_{i\in S} d_i (y_i - \bx_i^\top \bbeta_g^{(0)}) + o_p(N^{-1/2}). \label{eq:oracle-expansion}
\end{equation}
Therefore, $\sqrt{N}\bigl(\widehat{\mu}_{{\rm BC},\widehat{\pi}} - \mu\bigr) \xrightarrow{d} N(0,\sigma_{\rm BC}^2)$, where 
\begin{equation}\label{eq:asy-var}
  \sigma^2_{\rm BC} = 
  \mathbb{E} \bigg[ \frac{1 - \pi(\bX)}{\pi(\bX)} \{r^2(\bX) + \mathbb{V}(Y \mid \bX)\} \bigg].
\end{equation}
When the outcome regression is correctly specified, i.e.\ $m(\bx) = \bx^\top\boldsymbol{\beta}_0$ for some $\boldsymbol{\beta}_0$,
we have $r(\bx) = 0$ and~\eqref{eq:asy-var} reduces to $\sigma^2_{\rm BC} = \mathbb{E} \big\{ (\pi^{-1}(\bX) - 1) \mathbb{V}(Y \mid \bX) \big\}$.
\end{theorem}

%

\begin{corollary}
\label{cor:semi-eff}
Suppose the conditions of Theorem~\ref{thm:CLT-missing} hold and the
outcome regression model is correctly specified, $m(\bx) = \bx^\top
\boldsymbol{\beta}_0$. Then the asymptotic variance of the Bregman
calibration estimator,
\[
\sigma^2_{\rm BC}
= \mathbb{E}\!\left[\frac{1 - \pi(\bX)}{\pi(\bX)}\,
\mathbb{V}(Y \mid \bX)\right],
\]
attains the semiparametric efficiency bound for estimating the
finite-population mean $\mu = N^{-1} \sum_{i=1}^N y_i$ under the
non-probability sampling model with ignorable selection
$\delta \perp Y \mid \bX$.
\end{corollary}
 
 
The corollary shows that the Bregman calibration estimator attains the
semiparametric efficiency bound under correctly specified outcome regression,
matching the efficiency of the augmented inverse probability weighting
estimator of \citet{robins1994estimation} and the empirical balancing
calibration weighting estimator of \citet{chan2016globally}, but within a
broader generator class that admits efficiency tuning under
misspecification through the choice of $G(\cdot)$. Combined with the
design-optimality of the contrast-entropy generator under Poisson
sampling established in Section~\ref{sec:design}, this gives a unified
efficiency picture: the framework attains the semiparametric efficiency
bound when the outcome model is correctly specified, and the design-optimal
variance bound when the propensity follows a Poisson sampling design.

A consistent estimator of $\sigma^2_{\mathrm{BC}}$ is constructed
directly from the calibration residuals. Define
\begin{equation}
\label{eq:var-est-NPS}
\widehat\sigma^2_{\mathrm{BC}}
= \frac{1}{N} \sum_{i \in S}
  \widehat d_i (\widehat d_i - 1)
  \bigl(y_i - \bx_i^\top \widehat\bbeta_g^\star\bigr)^2,
\end{equation}
where $\widehat d_i = (\widehat\pi_i^{(-)})^{-1}$ and
$\widehat\bbeta_g^\star
= \bigl(\sum_{i \in S} \widehat q_i\, \bx_i \bx_i^\top\bigr)^{-1}
  \sum_{i \in S} \widehat q_i\, \bx_i y_i$
with $\widehat q_i = 1/g'(\widehat\omega_i)$.
The motivation for \eqref{eq:var-est-NPS} is the conditional identity
\begin{equation}
\label{eq:cond-id}
\mathbb{E}\!\bigl[\delta\, d (d - 1) (Y - \bX^\top \bbeta_g^{(0)})^2
   \,\big|\, \bX = \bx\bigr]
= \frac{1 - \pi(\bx)}{\pi(\bx)}
  \bigl\{\mathbb{V}(Y \mid \bx) + r^2(\bx)\bigr\},
\end{equation}
which follows from $\mathbb{E}(\delta \mid \bx) = \pi(\bx)$ and the
ignorable selection assumption $\delta \perp Y \mid \bX$, and which
yields
$\mathbb{E}[\delta\, d (d-1) (Y - \bX^\top \bbeta_g^{(0)})^2]
= \sigma^2_{\mathrm{BC}}$.

\begin{theorem}\label{thm:var-est-NPS}
Under the conditions in Lemma~\ref{lem:lambda-zero-est},
$\widehat\sigma^2_{\mathrm{BC}} \xrightarrow{p} \sigma^2_{\mathrm{BC}}$.
\end{theorem}

\section{Extension to high-dimensional auxiliary variables}
\label{sec:var_select}
 
When the auxiliary vector $\bx_i$ is high-dimensional relative to the sample
size $n = |S|$, the exact calibration constraint in~\eqref{bregman} may be
infeasible (e.g., when $p > n$) or may produce highly variable weights by
forcing exact balance on weakly relevant covariates. A natural extension
replaces exact balance with a soft tolerance constraint.
 
Let $\widetilde\bx_i \in \mathbb{R}^p$ denote the centered and scaled
non-intercept covariates, and let $\boldsymbol{\tau} = (\tau_1, \ldots,
\tau_p)^\top$ be a vector of positive tolerances. To preserve the correct
total scale, we impose exact calibration on the intercept,
\begin{equation}\label{eq:intercept}
\sum_{i \in S} \omega_i = n,
\end{equation}
and replace the remaining balance equations with the $\ell_q$ tolerance
constraint
\begin{equation}\label{eq:softbal}
N^{-1}\,\bigl\|\textstyle\sum_{i \in S}\omega_i\bigl(\widetilde
x_{i1}/\tau_1, \ldots, \widetilde x_{ip}/\tau_p\bigr)^\top\bigr\|_q \;\le\;
1,
\end{equation}
for some $1 \le q \le \infty$. The soft Bregman calibration (SBC) weights
are
\begin{equation}\label{eq:sbc}
\widehat\bomega_{\mathrm{SBC}}(\boldsymbol{\tau})
= \argmin_{\omega_i \in \mathcal V,\; i \in S}
\sum_{i \in S} D_G\bigl(\omega_i \,\|\, \omega_i^{(0)}\bigr)
\quad\text{subject to \eqref{eq:intercept} and \eqref{eq:softbal}},
\end{equation}
and the corresponding SBC estimator of the population mean is
$\widehat\mu_{\mathrm{SBC}}(\boldsymbol{\tau}) = n^{-1} \sum_{i \in S}
\widehat\omega_{\mathrm{SBC},i}(\boldsymbol{\tau})\, y_i$.
 
Lagrangian duality transforms this constrained problem into an unconstrained
problem with an explicit regularization structure.
 
\begin{proposition}[H\"older-regularized dual]\label{prop:holder-dual}
Let $1 \le q \le \infty$ and let $q^*$ denote its H\"older conjugate,
defined by $1/q + 1/q^* = 1$. Then the Lagrangian dual of~\eqref{eq:sbc} is
the unconstrained minimization
\begin{equation}\label{eq:holder-dual}
\min_{\lambda_0 \in \mathbb{R},\; \boldsymbol{\lambda} \in \mathbb{R}^p}
\Bigl[\,
\underbrace{\sum_{i \in S} F\bigl\{g(\omega_i^{(0)}) + \lambda_0
+ \widetilde\bx_i^\top \boldsymbol{\lambda}\bigr\} - n\lambda_0}
_{\text{smooth: Bregman fidelity}}
\;+\;
\underbrace{n\,\bigl\|(\tau_1\lambda_1, \ldots, \tau_p\lambda_p)\bigr\|_{q^*}}
_{\text{nonsmooth: H\"older penalty}}
\,\Bigr],
\end{equation}
and the SBC weights are recovered as $\widehat\omega_{\mathrm{SBC},i}(\boldsymbol{\tau})
= g^{-1}\bigl\{g(\omega_i^{(0)}) + \widehat\lambda_0(\boldsymbol{\tau})
+ \widetilde\bx_i^\top \widehat{\boldsymbol{\lambda}}(\boldsymbol{\tau})\bigr\}$.
\end{proposition}
 
The smooth term in~\eqref{eq:holder-dual} is governed by the convex
conjugate $F$ of the Bregman generator, while the nonsmooth penalty is
determined entirely by the primal constraint geometry $(q, \boldsymbol{\tau})$.
This decoupling means the generator and the regularization geometry can be
tuned separately. The case $q = \infty$ produces a weighted $\ell_1$ penalty
$n\sum_{k=1}^p \tau_k|\lambda_k|$, rendering many balance constraints
inactive and providing implicit, data-adaptive selection of calibration
variables, paralleling Lasso-type regularization \citep{wang2020b}; the case
$q = q^* = 2$ produces a weighted $\ell_2$ penalty analogous to ridge
regularization \citep{guggemos2010}. When outcome data are available, an
outcome-guided choice of tolerances $\tau_k = \tau / |\widehat\beta_k|$ for
a pilot estimator $\widehat{\boldsymbol{\beta}}$ provides a calibration
analogue of adaptive-Lasso weighting \citep{zou2006adaptive}; the detailed
construction and the cross-validation procedure for selecting the global
$\tau$ are given in Section D of the SM. Establishing
rates of convergence and oracle properties for the regularized estimator
remains an important direction for future research.

\section{Simulation Study}
\label{sec:sim}
\subsection{Simulation study one}
\label{subsec:sim1}
To investigate the double robustness of the proposed estimator, we performed a simulation study. For $i = 1, \cdots, N = 10,000$, $(\bm x_i, Y_i, \delta_i)$ are generated $500$ times repeatedly, where $\bx_i=(1, x_{i1}, x_{i2}, x_{i3}, x_{i4})^\top$. The following two outcome regression (OR) models were considered:
\begin{align*}
   \text{OR0: }& Y_i \;=\; 1 + x_{i1} - x_{i2} + e_i, \\ 
   \text{OR1: }& Y_i \;=\; 1 + x_{i1} - x_{i2} + x_{i1}x_{i2} + (x_{i2}^2-1) + e_i,
\end{align*}
along with the propensity score (PS) model
$\delta_i \sim \mbox{Bernoulli} (\pi_i)$, where $\pi_i = \exp ( \widetilde{c}_i) / \{1+ \exp (\widetilde{c}_i)\}$ with $\widetilde{c}_i$'s:
\begin{align*}
   \text{PS0: }& \widetilde{c}_i = 
-1 - 0.25 x_{i2} + 0.5  x_{i3},  \\ 
   \text{PS1: }& \widetilde{c}_i = -1 - 0.25(x_{i2} - 3)(x_{i3} - 4) + 0.5(x_{i2} - 2.5)^4,
\end{align*}
where $e_i \sim \mathcal{N}(0, 1)$, $x_{ij} \sim \text{TN}(2, 1, 0, 4)$ for $j = 1, \cdots, 4$ independently, and $\text{TN}(\mu, \sigma, a, b)$ denotes the normal distribution with mean $\mu$ and standard deviation $\sigma$ truncated to the interval $(a, b)$.
The auxiliary variable $\bx_i=(1, x_{i1}, \cdots, x_{i4})$ are observed for the whole population but $y_i$ are observed only when $\delta_i=1$.
We are interested in estimating $\mu=N^{-1}\sum_{i=1}^N y_i$ from the partially observed data. We consider four scenarios from the $2 \times 2$ combinations of PS and OR models. We consider three types of estimators of \(\mu\):
 
\begin{enumerate}
\item[\texttt{IPW}] {Inverse Probability Weighted} estimator:  \(\widehat{\mu}_{\rm IPW} = \big(\sum_{i\in S}\omega_i^{(0)}\big)^{-1}\sum_{i\in S}\omega_i^{(0)}y_i.\)

\item[\texttt{DS}] {Deville--S{\"a}rndal calibration:}
\(\widehat{\mu}_{\rm DS} = n^{-1}\sum_{i\in S}\widehat{\omega}_{{\rm DS},i}y_i,\)
where \(\widehat{\bomega}_{\rm DS}=(\widehat{\omega}_{{\rm DS},i};\, i\in S)\) minimizes
\(Q(\bomega\parallel\bomega^{(0)})\) in \eqref{deville} subject to $\sum_{i \in S} \omega_i \bm z_i =  n^{-1}\sum_{i=1}^N \bm z_i$.

\item[\texttt{BC}] {Bregman-divergence Calibration:}
\(\widehat{\mu}_{\rm BC} = n^{-1}\sum_{i\in S}\widehat{\omega}_{{\rm BC},i}y_i,\)
where \(\widehat{\bomega}_{\rm BC}=(\widehat{\omega}_{{\rm BC},i};\, i\in S)\) minimizes
\(D_G(\bomega\parallel\bomega^{(0)})\) in \eqref{bregman} subject to $\sum_{i \in S} \omega_i \bm z_i =  n^{-1}\sum_{i=1}^N \bm z_i$.
\end{enumerate}

We use the cross-fitted baseline weights $\omega_i^{(0)} = nN^{-1} /\widehat \pi_i^{(-)}$ in \eqref{eq:cf-initial-weights}, where $\widehat \pi_i^{(-)}$ is estimated from the logistic regression (\texttt{glm}) or logistic generalized additive model (\texttt{gam}) with penalized regression splines. The calibration constraints for \texttt{DS} and \texttt{BC} estimators use $\bm z_i= (1, x_{i1}, \cdots, x_{i4})^\top$. We consider three $G$ functions: exponential tilting (ET, $G( \omega) = \omega \log (\omega)$), empirical likelihood (EL, $G(\omega)=- \log (\omega)$), and Hellinger distance (HD, $G(\omega)= \big( \sqrt{\omega}-1 \big)^2$).

\begin{table}[!t]
\centering
\caption{Monte Carlo bias ($\times 10^2$), standard error ($\times 10^2$), and RMSE ($\times 10^2$) of calibration estimators across four PS/OR scenarios.
The \texttt{ET} row represents both \texttt{DS-ET} and \texttt{BC-ET}, which are numerically identical.}
\label{tab:sim1}
\renewcommand{\arraystretch}{1.05}
\setlength{\tabcolsep}{3.5pt}
\footnotesize
\begin{tabular}{@{}ll rrr rrr rrr rrr@{}}
\toprule
& & \multicolumn{3}{c}{PS0\,/\,OR0} & \multicolumn{3}{c}{PS0\,/\,OR1} & \multicolumn{3}{c}{PS1\,/\,OR0} & \multicolumn{3}{c}{PS1\,/\,OR1} \\
\cmidrule(lr){3-5} \cmidrule(lr){6-8} \cmidrule(lr){9-11} \cmidrule(lr){12-14}
$\widehat\pi$ & Method
  & Bias & SE & RMSE
  & Bias & SE & RMSE
  & Bias & SE & RMSE
  & Bias & SE & RMSE \\
\midrule
\multirow{6}{*}{\texttt{glm}}
  & \texttt{IPW}
  & $-0.0$ & 1.5 & 1.5
  & $-0.0$ & 3.6 & 3.6
  & $-24.0$ & 2.2 & 24.1
  & 180.4 & 9.4 & 180.6 \\
  & \texttt{ET}
  &  0.0 & 1.4 & 1.4
  & $-0.1$ & 2.2 & 2.2
  & $-0.0$ & 1.4 & 1.4
  &  56.4 & 2.5 & 56.4 \\
  & \texttt{DS-EL}
  &  0.0 & 1.4 & 1.4
  & $-0.1$ & 2.2 & 2.2
  & $-0.0$ & 1.4 & 1.4
  &  58.4 & 2.6 & 58.4 \\
  & \texttt{BC-EL}
  &  0.0 & 1.4 & 1.4
  & $-0.1$ & 2.2 & 2.2
  & $-0.0$ & 1.4 & 1.4
  &  49.8 & 2.4 & 49.8 \\
  & \texttt{DS-HD}
  &  0.0 & 1.4 & 1.4
  & $-0.1$ & 2.2 & 2.2
  & $-0.0$ & 1.4 & 1.4
  &  57.4 & 2.5 & 57.4 \\
  & \texttt{BC-HD}
  &  0.0 & 1.4 & 1.4
  & $-0.1$ & 2.2 & 2.2
  & $-0.0$ & 1.4 & 1.4
  &  52.9 & 2.4 & 52.9 \\
\addlinespace[4pt]
\hdashline
\multirow{6}{*}{\texttt{gam}}
  & \texttt{IPW}
  & $-0.1$ & 1.4 & 1.5
  &  0.1 & 3.3 & 3.3
  & $-0.8$ & 1.7 & 1.9
  &  4.2 & 2.7 & 5.0 \\
  & \texttt{ET}
  &  0.0 & 1.4 & 1.4
  & $-0.1$ & 2.0 & 2.0
  & $-0.0$ & 1.7 & 1.7
  &  0.2 & 2.0 & 2.0 \\
  & \texttt{DS-EL}
  &  0.0 & 1.4 & 1.4
  & $-0.1$ & 2.0 & 2.0
  & $-0.0$ & 1.7 & 1.7
  &  0.3 & 2.0 & 2.0 \\
  & \texttt{BC-EL}
  &  0.0 & 1.4 & 1.4
  & $-0.1$ & 2.0 & 2.0
  & $-0.0$ & 1.7 & 1.7
  &  0.2 & 2.0 & 2.0 \\
  & \texttt{DS-HD}
  &  0.0 & 1.4 & 1.4
  & $-0.1$ & 2.0 & 2.0
  & $-0.0$ & 1.7 & 1.7
  &  0.2 & 2.0 & 2.0 \\
  & \texttt{BC-HD}
  &  0.0 & 1.4 & 1.4
  & $-0.1$ & 2.0 & 2.0
  & $-0.0$ & 1.7 & 1.7
  &  0.2 & 2.0 & 2.0 \\
\bottomrule
\end{tabular}
\end{table}

\begin{figure}
    \centering
    \includegraphics[width=\linewidth]{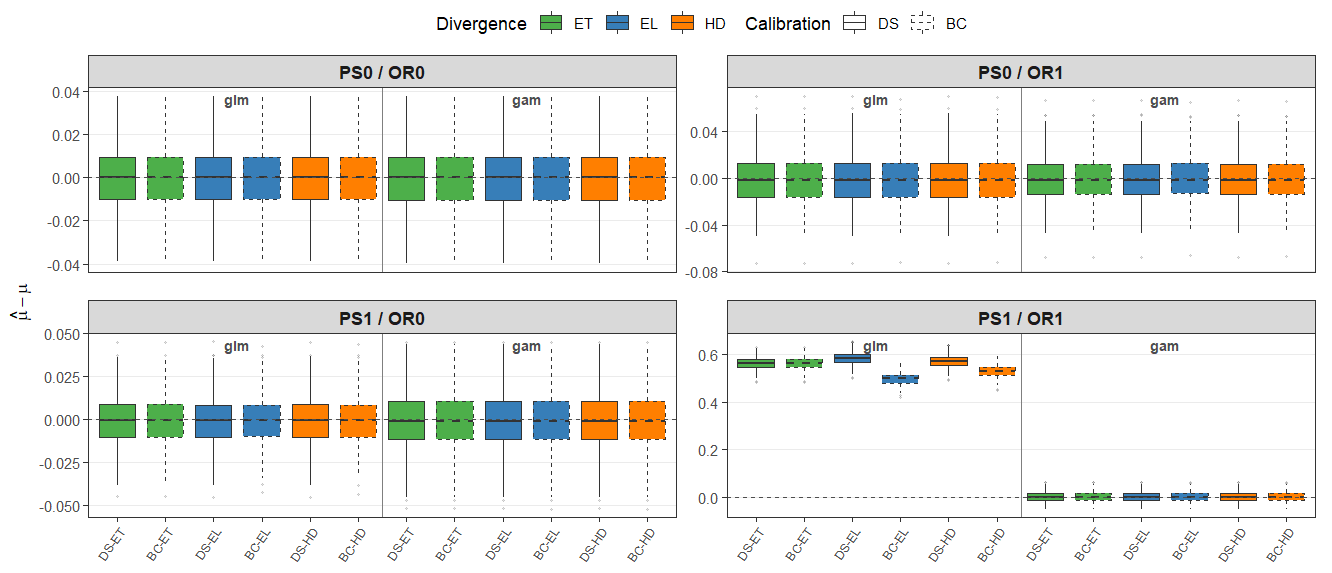}
    \caption{Boxplots for \texttt{DS} and \texttt{BC} estimators across four PS/OR scenarios with \texttt{glm} and \texttt{gam} propensity estimation.}
    \label{boxplots}
\end{figure}

Table~\ref{tab:sim1} and Figure~\ref{boxplots} present the Monte Carlo bias, standard error (SE), and RMSE of the calibration estimators across four PS/OR scenarios. The IPW estimator using \texttt{glm}-estimated propensity scores performs well under the correctly specified PS0 but suffers from substantial bias under PS1, while the \texttt{gam}-based IPW reduces bias relative to \texttt{glm} but retains nontrivial bias under PS1/OR1 due to the nonlinear propensity structure. All calibration estimators substantially reduce bias and RMSE relative to IPW. When at least one of the PS or OR models is correctly specified, the calibration estimators exhibit negligible bias, confirming double robustness. The \texttt{DS-ET} and \texttt{BC-ET} estimators are numerically identical across all configurations, consistent with the well-known equivalence of the two frameworks under the exponential tilting generator.

The most revealing contrasts appear in the doubly misspecified scenario PS1/OR1. With \texttt{glm} propensity estimation, all calibration estimators exhibit substantial bias, reflecting the inability of the linear logistic model to capture the highly nonlinear propensity structure in PS1. Under \texttt{glm} with PS1/OR1, the \texttt{BC} variants exhibit notably smaller bias than their \texttt{DS} counterparts, demonstrating the greater flexibility of the Bregman calibration objective. With \texttt{gam}-based propensity estimation under PS1/OR1, all calibration estimators exhibit negligible bias, indicating that the penalized spline model adequately captures the nonlinear propensity structure in PS1, and both \texttt{DS} and \texttt{BC} calibration successfully correct residual bias from the IPW step.

\subsection{Simulation study two}

In this simulation study, we evaluate the performance of the regularized Bregman calibration method introduced in Section~\ref{sec:var_select}. We generate a finite population of size $N=10,000$ from the OR0 and PS0 models in Section \ref{subsec:sim1}. We consider $\bm{X}_i = (X_{i1}, \ldots, X_{ip})^\top$ with $p = 500$, generated from $\bm{X}_i \sim \mathcal{N}(\bm{2}, \bm{\Sigma})$
where $\Sigma_{jk} = \rho^{|j-k|}$ with $\rho = 0.5$. All simulation steps are repeated over $B = 500$ Monte Carlo replications. 

For each replication, we apply the soft Bregman calibration method. We use the cross-fitted baseline weights $\omega_i^{(0)} = nN^{-1} /\widehat \pi_i^{(-)}$ in \eqref{eq:cf-initial-weights}, where $\widehat \pi_i^{(-)}$ is estimated from logistic regression refitted on the LASSO-selected variables. We consider three choices of Bregman divergence: \texttt{ET}, \texttt{EL}, and \texttt{HD}. The outcome-guided penalty weights $\tau_k = \tau / |\widehat\beta_k|$ are constructed from two pilot estimators: an OLS fit of the outcome model on the sampled data, and a LASSO-refitted (post-LASSO OLS) estimator. Based on this setup, we consider the following calibration estimators:
\begin{itemize}
\item[\texttt{Full}] The full calibration estimator using all available covariates.
\item[\texttt{Oracle}] The oracle calibration estimator using only the covariates in the outcome regression model, namely $(1, X_1, X_2)$.
\item[\texttt{SBC}] The soft calibration estimator in \eqref{eq:softbal}, using $\ell_q$ norms and a fixed $\tau$. We use $q = 1, 2, \infty$.
\end{itemize}

\begin{table}[!t]
\centering
\caption{Monte Carlo bias ($\times 10^2$), standard error
  ($\times 10^2$), and RMSE ($\times 10^2$) of the calibration
  estimators in simulation study two.
  For \texttt{SBC}, results are reported at fixed
  $\tau = 5 \times 10^{-4}$.
  Panel~(a) uses OLS pilot coefficients; Panel~(b) uses
  LASSO-refitted pilot coefficients.}
\label{tab:sim2}
\renewcommand{\arraystretch}{1.1}
\setlength{\tabcolsep}{3.5pt}
\footnotesize
\begin{tabular}{@{}ll rrr rrr rrr@{}}
\toprule
& & \multicolumn{3}{c}{\texttt{EL}}
  & \multicolumn{3}{c}{\texttt{ET}}
  & \multicolumn{3}{c}{\texttt{HD}} \\
\cmidrule(lr){3-5} \cmidrule(lr){6-8} \cmidrule(lr){9-11}
Estimator & $q$
  & Bias & SE & RMSE
  & Bias & SE & RMSE
  & Bias & SE & RMSE \\
\midrule
\texttt{IPW} & ---
  & \multicolumn{9}{c}{Bias $= 0.528$, \quad SE $= 1.975$, \quad RMSE $= 2.043$} \\[3pt]
\texttt{Full} & ---
  & $-0.003$ & 1.532 & 1.530
  & $0.004$ & 1.498 & 1.497
  & $0.001$ & 1.512 & 1.510 \\
\texttt{Oracle} & ---
  & $-0.020$ & 1.415 & 1.414
  & $-0.021$ & 1.415 & 1.414
  & $-0.021$ & 1.415 & 1.414 \\[3pt]
\midrule
\multicolumn{11}{@{}l}{\textit{(a) OLS pilot}} \\[3pt]
\texttt{SBC}
  & $q=1$
  &  0.003 & 1.515 & 1.513
  &  0.005 & 1.490 & 1.489
  &  0.004 & 1.500 & 1.498 \\
  & $q=2$
  &  0.005 & 1.491 & 1.489
  &  0.005 & 1.476 & 1.474
  &  0.004 & 1.481 & 1.480 \\
  & $q=\infty$
  & $-0.011$ & 1.419 & 1.418
  & $-0.010$ & 1.418 & 1.417
  & $-0.011$ & 1.418 & 1.417 \\
\midrule
\multicolumn{11}{@{}l}{\textit{(b) LASSO-refit pilot}} \\[3pt]
\texttt{SBC}
  & $q=1$
  & $-0.012$ & 1.417 & 1.416
  & $-0.013$ & 1.417 & 1.416
  & $-0.012$ & 1.417 & 1.416 \\
  & $q=2$
  & $-0.012$ & 1.418 & 1.416
  & $-0.013$ & 1.417 & 1.416
  & $-0.012$ & 1.417 & 1.416 \\
  & $q=\infty$
  & $-0.011$ & 1.420 & 1.419
  & $-0.011$ & 1.419 & 1.418
  & $-0.011$ & 1.420 & 1.418 \\
\bottomrule
\end{tabular}
\end{table}

\begin{figure}[!t]
    \centering
    \includegraphics[width=\linewidth]{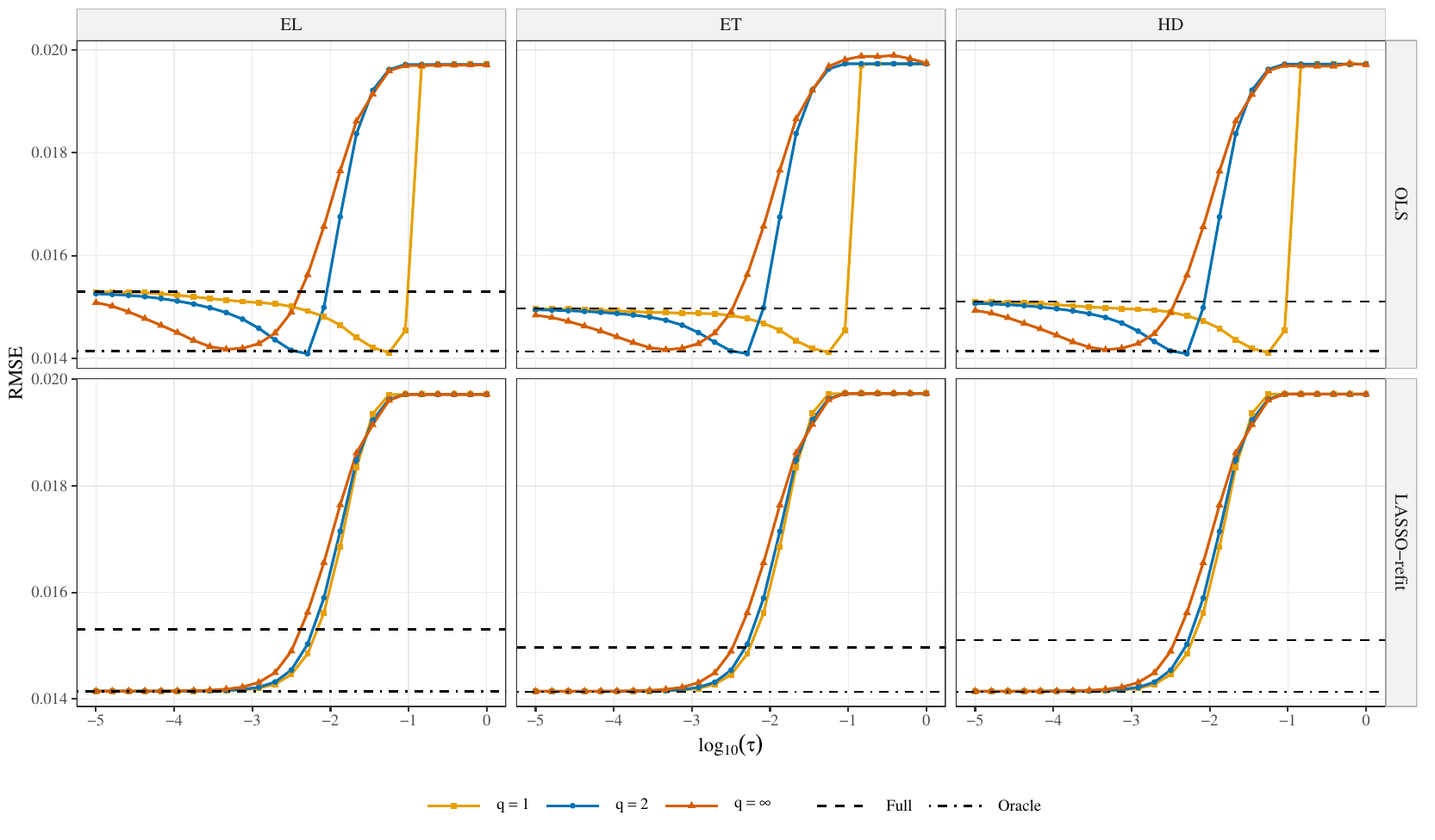}
    \caption{RMSE ($\times 10^{2}$) of the \texttt{SBC}
      estimator as a function of $\log_{10}(\tau)$.
      Top row: OLS pilot; bottom row: LASSO-refit pilot.
      Columns correspond to the three divergence functions.
      Dashed and dot-dashed lines indicate the \texttt{Full} and
      \texttt{Oracle} baselines, respectively.}
    \label{fig:sim2}
\end{figure}

Table~\ref{tab:sim2} and Figure~\ref{fig:sim2} present the results. The \texttt{IPW} estimator exhibits the largest RMSE, while \texttt{Full} and \texttt{Oracle} achieve substantially lower RMSE with negligible bias; the gap between the two reflects the cost of calibrating against $p = 500$ superfluous covariates. The choice of divergence function has virtually no effect on performance for any estimator. The behavior of \texttt{SBC} depends strongly on the pilot estimator. With OLS pilot coefficients (Panel~a), only $q = \infty$ matches the \texttt{Oracle}, because the $\ell_\infty$-norm acts as a coordinate-wise box constraint that concentrates calibration on covariates with large $|\widehat\beta_k|$; $q = 1$ and $q = 2$ distribute the penalty broadly and remain closer to \texttt{Full}. With LASSO-refitted pilot coefficients (Panel~b), all three $q$-norms achieve near-\texttt{Oracle} performance, because the sparse pilot $\widehat\beta$ effectively removes irrelevant covariates from the calibration regardless of $q$.

Figure~\ref{fig:sim2} shows how the RMSE of \texttt{SBC} varies with $\tau$. Under OLS, the $q = \infty$ curve is U-shaped, entering the \texttt{Full}--\texttt{Oracle} band for $\tau \in [10^{-4}, 10^{-3}]$ with an uptick at smaller $\tau$ due to near-exact constraint difficulty. Under the LASSO-refit pilot, all three $q$-norms yield overlapping S-shaped curves that descend monotonically into the \texttt{Full}--\texttt{Oracle} band without such an uptick, since the sparse pilot keeps the effective number of calibration constraints small. In both cases, increasing $\tau$ beyond $10^{-2}$ reverts the estimator toward \texttt{IPW}, and the patterns are consistent across all divergence functions.

\section{Real data analysis}
\label{sec:realdata}

We use Large Pelagics Intercept Survey (LPIS) 
at  National Oceanic and Atmospheric Administration (NOAA)
as a test bed for calibration weighting in a setting where inclusion probabilities vary and rich frame-level auxiliary information is available \citep{foster2008large}. The public LPIS sampling frame contains 74{,}253 site--day--time units (saltwater fishing sites crossed with day and time block) across nine Atlantic states during June--October. LPIS employs a complex stratified, multi-stage design, with unequal-probability sampling without replacement (PPSWOR). Selection probabilities are driven by cluster-level fishing pressure and day type (e.g., weekday vs. weekend; tournament vs. non-tournament), and field assignments vary in duration (about 2--8 hours). In practice, operational flexibilities (e.g., variable site choice within clusters and emphasis on afternoon sampling) can induce coverage gaps that are difficult to fully account for if major design features are ignored in estimation, motivating careful use of weighting and calibration.

Treating the sampling frame as a finite population, we generate population-level boat-trip counts as independent (truncated) Poisson draws and simulate eleven catch outcomes under alternative trip--catch relationships (ranging from no association/binary catch to retention and harvest mechanisms under moderate/high catch rates). The trip counts and eleven catch outcomes (12 response variables total) are treated as $y_i$. Calibration covariates are available at the frame level and include fishing pressure, day type (weekend versus weekday), boat mode (charter versus private), county, and a noise variable included as a negative control. 

For 1{,}000 repeated replications, we select units under unequal--probability sampling with
expected sample size $n=1{,}135$, using selection probabilities proportional to
frame-level inclusion probabilities. We compare the Deville--S\"arndal (DS) and
Bregman-divergence calibration (BC) estimators under exponential tilting (ET), empirical likelihood (EL), and Hellinger distance (HD); we also include BC with contrast entropy (CE). Performance is summarized across the 12 outcomes, and we report RMSE scaled
relative to BC--CE to facilitate comparisons. 

\begin{figure}[!t]
  \centering
  \includegraphics[width=0.9\linewidth]{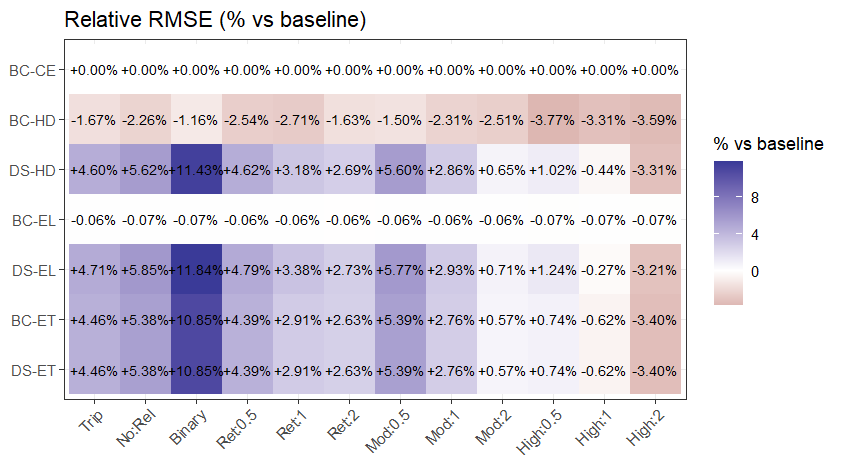}
  \caption{LPIS unequal--probability sampling: heatmap of scaled relative RMSE across 12 outcomes. Values below 0 indicate improved efficiency relative to BC--CE.}
  \label{fig:lpis_rmse}
\end{figure}

Under unequal--probability sampling, the estimators are design-unbiased but show meaningful differences in efficiency across study variables. Figure~\ref{fig:lpis_rmse} summarizes relative RMSE, $100 \times \big(\frac{\text{RMSE(method)}}{\text{RMSE(BC-CE)}} - 1\big)$, over the 12 outcomes for a range of calibration covariate specifications. BC--HD attains the
lowest RMSE in most settings. DS--ET and BC--ET coincide, DS--HD is broadly
comparable to DS--ET, and EL-based methods are typically slightly less efficient in this example.

This study illustrates how the Bregman framework encompasses classical calibration while allowing alternative distance choices that can improve efficiency when inclusion probabilities vary. Although \texttt{BC-CE} is optimal under Poisson sampling in theory, the LPIS emulation reflects additional complexities, under which \texttt{BC-HD} performs comparably or better.

\section{Discussion}
\label{sec:discussion}
We have proposed a unified calibration weighting framework rooted in the
Bregman divergence.  By formulating calibration directly in the weight space
rather than on weight ratios, the framework reveals calibration as a geometric
projection with a primal--dual symmetry: both the weight-space and
multiplier-space problems are instances of Bregman divergence minimization, the
latter reducing a high-dimensional constrained problem to an unconstrained
optimization in $p$ dimensions.

A central finding is that, unlike the Deville--S\"arndal framework, the regression coefficient in
the equivalent debiased prediction estimator depends explicitly on the choice of
the generator~$G(\cdot)$.  This dependence is a key strength: it enables
deliberate efficiency tuning through generator selection.  The contrast-entropy
function satisfies the condition for
design-optimality under Poisson sampling.  When propensity scores must be estimated, the
cross-fitting procedure combined with calibration yields doubly robust
estimation, requiring only that the product of the propensity estimation error
and the outcome approximation error vanishes faster than $n^{-1/2}$.

We further developed a regularized extension that replaces exact balance with $\ell_q$-norm tolerance constraints, with Lagrangian duality yielding a Hölder-conjugate dual penalty. Establishing rates of convergence and oracle properties for this regularized estimator is an important topic for future research.


\section*{Data Availability Statement}

The replication code, simulation results, and data supporting the findings of this study will be openly available in Zenodo at 
https://zenodo.org/records/19362709.

\section*{Acknowledgements}

The authors would like to thank Jay Breidt and Chien-Min Huang (NORC at the University of Chicago), John Foster and Yong-Woo Lee (NOAA Fisheries), and Anthony Kaufman and Daemian Schreiber (NOAA affiliates) for providing the simulated LPIS population used in the real data analysis.

\bibliographystyle{apalike}
\bibliography{ref}

@string{jasa = "Journal of the American Statistical Association"}

@article{montanari1987,
  title = {Post-sampling efficient {QR}-prediction in large-sample surveys}, 
   author = {Montanari, G.~E.},
    year = {1987},
    journal = {International Statistical Review},  
    volume = {55},
    Number = {2},
    pages = {191--202}
}

@article{hainmueller2012entropy,
  title={Entropy balancing for causal effects: A multivariate reweighting method to produce balanced samples in observational studies},
  author={Hainmueller, Jens},
  journal={Political analysis},
  volume={20},
  number={1},
  pages={25--46},
  year={2012},
  publisher={Cambridge University Press}
}

@article{wurao2006, 
   title = {Pseudo Empirical Likelihood Ratio Confidence Intervals for Complex Surveys}, 
    year = {2006}, 
    author = {Wu, C and Rao, J.~N.~K.}, 
     journal = {Canadian Journal of Statistics},
     volume = {34}, 
  number={3},
      pages = {359--375}
}

@article{deville1992,
  title={Calibration estimators in survey sampling},
  author={Deville, Jean-Claude and S{\"a}rndal, Carl-Erik},
  journal={Journal of the American statistical Association},
  volume={87},
  number={418},
  pages={376--382},
  year={1992},
  publisher={Taylor \& Francis}
}

@article{devaud2019,
   title = {Deville and {S}{\"a}rndal's calibration: revisiting a 25-years-old successful optimization problem (with discussion)}, 
   author = {Devaud, D. and Till{\'e}, Y.}, 
   year = {2019}, 
   journal = {Test}, 
   volume = {28}, 
    pages = {1033--1065}
}

@book{csiszar2004, 
   author = {Csisz{\'a}r, Imre and Shields, P.~C.}, 
   year = {2004},  
   title = {Information theory and Statistics: A tutorial}, 
   publisher = {Now Publishers Inc.} 
}

@article{zhao2019,
     author = {Zhao, Q.},
     title = {Covariate balancing propensity score by Tailored loss functions}, 
     year = {2019}, 
     journal = {The Annals of Statistics}, 
     volume = 47,
     pages = {965--993} 
}

@article{chan2016globally, 
    author = {Chan, Kwun Chuen Gary and Yam, Sheung Chi Phillip and Zhang, Zheng},
    title = {Globally Efficient Non-Parametric Inference of Average Treatment Effects by Empirical Balancing Calibration Weighting},
    journal = {Journal of the Royal Statistical Society Series B: Statistical Methodology},
    volume = {78},
    number = {3},
    pages = {673-700},
    year = {2016},
    doi = {10.1111/rssb.12129},
    url = {https://doi.org/10.1111/rssb.12129},
    eprint = {https://academic.oup.com/jrsssb/article-pdf/78/3/673/49236400/jrsssb\_78\_3\_673.pdf},
}

@article{robins1994estimation,
  title={Estimation of regression coefficients when some regressors are not always observed},
  author={Robins, James M and Rotnitzky, Andrea and Zhao, Lue Ping},
  journal={Journal of the American statistical Association},
  volume={89},
  number={427},
  pages={846--866},
  year={1994},
  publisher={Taylor \& Francis}
}

@article{hainmueller2012,
  title={Entropy balancing for causal effects: A multivariate reweighting method to produce balanced samples in observational studies},
  author={Hainmueller, Jens},
  journal={Political Analysis},
  volume = {20},
  pages={25--46},
  year={2012},
  publisher={JSTOR}
}

@article{gneiting2007strictly,
  title={Strictly proper scoring rules, prediction, and estimation},
  author={Gneiting, Tilmann and Raftery, Adrian E},
  journal={Journal of the American statistical Association},
  volume={102},
  number={477},
  pages={359--378},
  year={2007},
  publisher={Taylor \& Francis}
}

@article{imai2014,
           author = {Imai, K. and Ratkovic, M. }, 
           year = {2014},
           journal = {Journal of the Royal Statistical Society: Series B},
           title = {Covariate balancing propensity score}, 
           volume = {76}, 
           pages = {243--263} 
}

@article{breidt2017model,
  title={Model-Assisted Survey Estimation with Modern Prediction Techniques},
  author={Breidt, F Jay and Opsomer, Jean D},
  journal={Statistical Science},
  volume={32},
  number={2},
  pages={190--205},
  year={2017},
  publisher={Institute of Mathematical Statistics}
}

@article{robinson1983asymptotic,
  title={Asymptotic properties of the generalized regression estimator in probability sampling},
  author={Robinson, PM and S{\"a}rndal, Carl Erik},
  journal={Sankhy{\=a}: The Indian Journal of Statistics, Series B},
  pages={240--248},
  year={1983},
  publisher={JSTOR}
}

@article{wang2020b, 
  title = {Minimal dispersion approximately balancing weights: asymptotic properties and practical considerations}, 
  year = {2020}, 
  author = {Wang, Y. and Zubizarreta, J.~R.}, 
  journal = {Biometrika}, 
  volume = {107}, 
  pages = {93--105} 
}

@article{guggemos2010,
  title={Penalized calibration in survey sampling: Design-based estimation assisted by mixed models},
  author={Guggemos, Fabien and Till{\'e}, Yves},
  journal={Journal of statistical planning and inference},
  volume={140},
  pages={3199--3212},
  year={2010},
  publisher={Elsevier}
}

@article{kwon2024, 
    title = { Debiased calibration estimation using generalized entropy  in survey sampling }, 
    author = {Kwon, Y. and Kim, J.~K. and Qiu, Y.}, 
     year = {2025},
  journal = jasa, 
     note = {
https://doi.org/10.1080/01621459.2025.2537452  
     }
}

@article{chen2020doubly,
  title={Doubly robust inference with nonprobability survey samples},
  author={Chen, Yilin and Li, Pengfei and Wu, Changbao},
  journal={Journal of the American Statistical Association},
  volume={115},
  number={532},
  pages={2011--2021},
  year={2020}
}

@book{amari2000, 
   title={Methods of information geometry},
   author={Amari, Shun-ichi and Nagaoka, Hiroshi}, 
    year = {2000}, 
   publisher = {American Mathematical Society}
}

@article{fuller2002regression,
  title={Regression estimation for survey samples},
  author={Fuller, Wayne A},
  journal={Survey Methodology},
  volume={28},
  number={1},
  pages={5--24},
  year={2002}
}

@article{foster2008large,
  title={Large pelagics survey: methodology overview and issues},
  author={Foster, J and Salz, R and Sminkey, TR and Van Voorhees, D and Andrews, R and Lai, HL},
  journal={ICES (International Council for the Exploration of the Sea) CM},
  year={2008}
}

@article{zou2006adaptive,
  title={The adaptive lasso and its oracle properties},
  author={Zou, Hui},
  journal={Journal of the American Statistical Association},
  volume={101},
  number={476},
  pages={1418--1429},
  year={2006},
  publisher={Taylor \& Francis}
}

@article{wu2022statistical,
  title={Statistical inference with non-probability survey samples},
  author={Wu, Changbao},
  journal={Survey Methodology},
  volume={48},
  number = {2},
  pages={283--311},
  year={2022}
}

\end{document}